\documentclass[aps,twoside,superscriptaddress,twocolumn,floatfix,a4paper,prl,longbibliography]{revtex4-2}
\usepackage{color}
\usepackage{graphicx}
\usepackage[utf8x]{inputenc}
\usepackage[T1]{fontenc}
\usepackage{siunitx}
\usepackage{amstext}
\usepackage{textgreek}
\usepackage{epstopdf}
\usepackage{amsfonts}
\usepackage{amsmath}
\usepackage{multirow}

\usepackage{marvosym}
\usepackage[dvipsnames]{xcolor}

\begin{document}

\title{Sensitivity versus selectivity in entanglement detection via collective witnesses}

\author{Vojtěch Tr\'{a}vn\'{i}\v{c}ek}
\email{travnicekv@fzu.cz}
\affiliation{Institute of Physics of the Czech Academy of Sciences, Joint Laboratory of Optics of PU and IP AS CR, 17. listopadu 50A, 772 07 Olomouc, Czech Republic}

\author{Jan Roik}
%\email{jan.roik@upol.cz}
\affiliation{Joint Laboratory of Optics of Palacký University and Institute of Physics of Czech Academy of Sciences, 17. listopadu 12, 771 46 Olomouc, Czech Republic}

\author{Karol Bartkiewicz} 
%\email{bark@amu.edu.pl}
\affiliation{Faculty of Physics, Adam Mickiewicz University, PL-61-614 Pozna\'n, Poland}

\author{Antonín \v{C}ernoch} 
%\email{acernoch@fzu.cz}
\affiliation{Institute of Physics of the Czech Academy of Sciences, Joint Laboratory of Optics of PU and IP AS CR, 17. listopadu 50A, 772 07 Olomouc, Czech Republic}

%\author{Tomáš Fürst}
%\email{tomas.furst@upol.cz}
%\affiliation{Faculty of Science Department of Mathematical Analysis and Applications of Mathematics Palacký University, 771 46 Olomouc, Czech Republic}

\author{Paweł Horodecki}
%\email{pawel.horodecki@pg.edu.pl}
\affiliation{Faculty of Applied Physics and Mathematics Gda\'nsk University of Technology, 80-233 Gda\'nsk, Poland}
\affiliation{International Center for Theory of Quantum Technologies University of Gda\'nsk, Wita Stwosza 63, 80-308 Gda\'nsk, Poland}

\author{Karel Lemr}
\email{k.lemr@upol.cz}
\affiliation{Joint Laboratory of Optics of Palacký University and Institute of Physics of Czech Academy of Sciences, 17. listopadu 12, 771 46 Olomouc, Czech Republic}

\begin{abstract}
In this paper, we present a supervised learning technique that utilizes artificial neural networks to design new collective entanglement witnesses for two-qubit and qubit-qutrit systems. Machine-designed collective entanglement witnesses allow for continuous tuning of their sensitivity and selectivity. These witnesses are, thus, a conceptually novel instrument allowing to study the sensitivity vs. selectivity trade-off in entanglement detection. The chosen approach is also favored due to its high generality, lower number of required measurements compared to quantum tomography, and potential for superior performance with regards to other types of entanglement witnesses. Our findings could pave the way for the development of more efficient and accurate entanglement detection methods in complex quantum systems, especially considering realistic experimental imperfections.
\end{abstract}

\date{\today}

\maketitle
\paragraph*{Introduction.}
Quantum entanglement is a fascinating phenomenon that has intrigued physicists ever since it was predicted in 1935 \cite{bib:Einstein:EPR}. In recent decades, quantum entanglement has been recognized as an essential resource for various quantum information processing tasks \cite{bib:Zeilinger:QT,bib:Pan:swapp,bib:e91}, providing quantum computational and security superiority over classical information science \cite{bib:Harrow:Qsupremacy,bib:Zhang:QKD,zurek1982single,bib:Ren:ANN_QC}. Although entanglement is a straightforward consequence of the principle of superposition applied to a multipartite quantum state, there is still an open problem in finding a robust and efficient method for detecting entanglement in general quantum systems \cite{bib:hiesmayr:openproblem,bib:Gharibian:openproblem}. The holy grail of entanglement detection would be an approach that minimizes the number of required measurements while maximizing both sensitivity and selectivity. Achieving this balance is, however, challenging because increasing the sensitivity of detection usually leads to a higher false positive rate, while increasing the selectivity may result in a higher false negative rate.

\begin{figure}[t!]
\includegraphics[width=8.5cm]{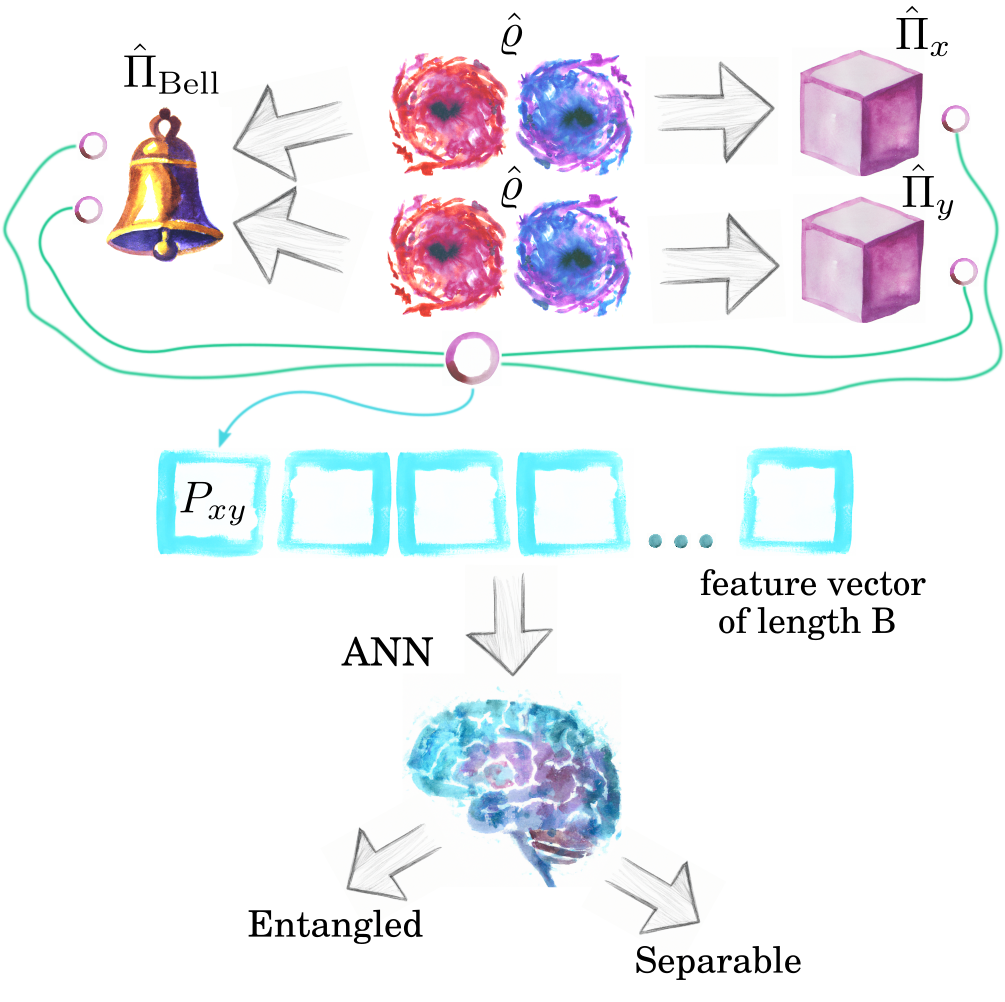}
\caption{\label{fig:scheme} The conceptual diagram of a collective measurement process and subsequent decision making by artificial neural network (ANN) model of collective entanglement witness (CEW): two instances of the system under investigation, denoted as $\hat{\varrho}$, are measured simultaneously. During the measurement, one subsystem from each instance undergoes a local projection using operators $\hat{\Pi}_{x}$ and $\hat{\Pi}_{y}$, while the other two subsystems are non-locally projected onto a Bell state. The results of these measurements, denoted as $P_{xy}$, are used to create a feature vector that serves as input for an ANN, which then according to some predetermined threshold labels the investigated system as entangled or separable.}
\end{figure}

Currently, there exist several analytical methods to detect and in some cases also quantify entanglement, particularly in simpler systems. In principle, one can perform a full quantum state tomography followed by density matrix estimation \cite{bib:Jezek:Qtom,bib:Miranowicz:Qtom}. Using the density matrix, one can determine criterion such as the Negativity and, thus, differentiate between separable and entangled systems \cite{bib:Peres:PPT,bib:Zyczkowski:negativity}. The problem with this approach is that the number of measurements required for a full quantum state tomography increases exponentially as the system under investigation becomes more complex. Researchers have suggested a way to tackle this problem by introducing a direct method that employs entanglement witnesses \cite{bib:Horodecki:EW,bib:Guhne:EntDet}. Entanglement witnesses are certain quantities that can be calculated directly by performing various measurements on the system. If the measurement outcomes fall within a specific range, it indicates the presence of entanglement in the system. These include e.g., the famous Clauser-Horne-Shimony-Holt (CHSH) inequality  \cite{bib:clauser:CHSH}, full entangled fraction (FEF) \cite{bib:bennett:FEF,bib:bartkiewicz:FEF,bib:Miranowicz:FEF} or nonlinear entropic witness (EW) \cite{bib:castro:entropy_W,bib:vedral:entropy_W}. The most measurement efficient class of entanglement witnesses are the collective witnesses (CEWs), which necessitates performing measurements collectively on multiple copies of the quantum system under investigation \cite{bib:Mintert:CEW,bib:Mintert:concurrence}. An example of this type of entanglement witness is the Collectibility \cite{bib:rudnicki:CEW,bib:Rudnicki:collectibility}. A conceptual diagram of a collective measurement process is shown in Fig.~\ref{fig:scheme}.

Analytical entanglement witnesses are designed to be perfectly selective, i.e., to never classify a separable state as entangled \cite{bib:Horodecki:EW,bib:terhal:EW}. While this property is advantageous in certain scenarios, it also considerably reduces the sensitivity of the entanglement witnesses, i.e., the ability to correctly identify many entangled states \cite{bib:horodecki:QE,bib:Bae:ent}. Note that in experimental reality, even analytical entanglement witnesses are prone to false positive classification due to unavoidable measurement noise. This trade-off between selectivity and sensitivity raises the question of whether it is possible to improve the sensitivity of entanglement witnesses at only a slight expense of their selectivity. We provide the answer in this paper where we present the application of a supervised learning technique that utilizes artificial neural networks (ANNs) to design new collective entanglement witnesses for two-qubit and qubit-qutrit systems. There are several key features this method brings to the area of entanglement detection. Firstly, machine designed CEWs in contrast to analytical witnesses allow for continues tuning of their sensitivity and selectivity. Secondly, the chosen CEW approach is favored due to its high generality \cite{bib:Augusiak:UnivesalEW}, lower number of required measurements compared to quantum tomography \cite{bib:Bovino:QE}, and its potential for superior performance with regards to other types of entanglement witnesses \cite{bib:horodecki:QE}. Thirdly, entanglement witnesses designed by ANNs optimize the sensitivity-selectivity trade-off (ROC curves) given arbitrary amount of input information e.g., number of measurements. To our best knowledge, the analysis of the sensitivity-selectivity trade-off, as performed in this study, has not been previously explored in the context of entanglement detection.

%====================================

\paragraph*{Methods.}

CEWs are nonlinear functionals of a density matrix $\hat{\varrho}$ and rely on joint measurements of multiple copies of a system under investigation to detect entanglement \cite{bib:horodecki:QE}. Our focus is on CEWs that require simultaneous measurements on two copies of the system, as this type of CEWs is more experimentally feasible \cite{bib:Bovino:QE,bib:schmidt:CEW,bib:lemr:CEW}. Additionally, the experimental layout of a two-copy CEWs is similar to the geometry of the entanglement swapping protocol \cite{bib:Travnicek:ES}, which is a core principle of quantum repeaters and relays – essential devices for quantum communications \cite{bib:Briegel:repeater,bib:Jacobs:relay}.

To perform collective measurements on a two-qubit (or qubit-qutrit) system, represented by a density matrix $\hat{\varrho}$, it is necessary to prepare two instances of this system. These instances are then described as $\hat{\varrho}_{T} = \hat{\varrho} \otimes \hat{\varrho}$. From each instance, one subsystem is selected for joint nonlocal Bell-state projection, while the remaining subsystem is projected locally (see Fig.~\ref{fig:scheme}). In the case of qubit-qutrit system, it is the qubits that undergo the joint nonlocal projection. The outcome of a collective measurement is the probability of successfully projecting onto a singlet Bell state, given a particular pair of local projections. This is represented by the equation

\begin{equation}
P_{xy} = \frac{\mathrm{Tr}[(\hat{\varrho}_{T})(\hat{\Pi}_{x}\otimes\hat{\Pi}_{\mathrm{Bell}}\otimes\hat{\Pi}_{y})]}{\mathrm{Tr}[(\hat{\varrho}_{T})(\hat{\Pi}_{x}\otimes\hat{1}\otimes\hat{\Pi}_{y})]},
\end{equation}
where $\hat{\Pi}_{x(y)}$ represents a projection onto a single qubit (or qutrit) state, $\hat{\Pi}_{\mathrm{Bell}}$ is a two-qubit projection onto singlet Bell state, and $\hat{1}$ denotes four- (or six-) dimensional identity matrix. Each unique combination of single qubit (or qutrit) projections $\hat{\Pi}_{x}$ and $\hat{\Pi}_{y}$ represents one measurement configuration $P_{xy}$. We use these measurement outcomes to train our ANN models in predicting Negativity, an extension of the Peres-Horodecki criterion (PPT) \cite{bib:Peres:PPT} and a measure of quantum entanglement. The analytical Negativity is defined as the absolute sum of negative eigenvalues $\lambda_{i}$ of a partial transposition of $\hat{\varrho}$
\begin{equation}
\mathcal{N}(\hat{\varrho}) = \Bigg |\sum_{\lambda_{i}<0}\lambda_{i}\Bigg |.
\end{equation}

The training, testing and validation datasets consist of feature vectors of length $B$, where each vector corresponds to a randomly sampled density matrix $\hat{\varrho}$, and $B$ represents the number of realized measurement configurations $P_{xy}$ upon the density matrix $\hat{\varrho}_{T}$. It is important to highlight that there are various methods available for generating uniformly sampled density matrices \cite{bib:maziero:sampling,bib:Zyczkowski:negativity}. These methods differ in terms of the frequency at which separable and entangled systems occur within the generated ensemble. In our study, we employed a method described in the Ref.~\cite{bib:Zyczkowski:negativity}, which yields a prevalence of $0.63$ for separable two-qubit systems and $0.38$ for separable qubit-qutrit systems. However, to ensure that our ANN models were trained on unbiased ensembles, we intentionally adjusted the prevalence to $0.5$. For more details on random density matrix generation see the Appendix. The training dataset contained $4\times 10^5$ feature vectors, while an additional independent testing dataset consisted of $1\times 10^5$ feature vectors and was utilized to assess the performance of the models. Negativity was calculated analytically for a given density matrix $\hat{\varrho}$ and used to label the matrix as entangled or separable. To prevent over-fitting, a cross-validation strategy was implemented during the training process using a validation dataset of $1\times 10^5$ feature vectors. An early stopping condition was activated when the ANNs intermediate model performance on the cross-validation set did not demonstrate improvement in a monitored quantity for five consecutive epochs. The optimal balance between the model's complexity and precision was achieved when the ANN was composed of two hidden layers containing 32 and 16 nodes, respectively. Increasing the number of nodes beyond these numbers did not result in any notable improvement, whereas decreasing the number of nodes by half had a negative impact on the model's precision.

To evaluate the performance of machine designed CEWs in terms of both sensitivity and selectivity, it is useful to consider the receiver operating characteristic (ROC) curve, which plots the true positive rate (TPR) against the false positive rate (FPR) for different decision thresholds of the CEW model. In our assessment we compared the predicted Negativity of the system with a certain threshold value. Systems with predicted Negativity below the threshold were classified as entangled, while those exceeding the threshold were classified as separable. By comparing the number of correctly classified entangled systems to the actual number of entangled systems in the ensemble, we obtained the true positive rate. Similarly, we determined the false positive rate by comparing the number of incorrectly classified separable systems to the total number of separable systems.

%====================================
\paragraph*{Results.}

In this study, we have utilized supervised learning techniques and artificial neural networks to develop new collective entanglement witnesses for two-qubit and qubit-qutrit systems. The primary goal of this work is to enhance the sensitivity of these witnesses at slight expanse of their selectivity.

\begin{figure}[t!]
\includegraphics[width=8.5cm]{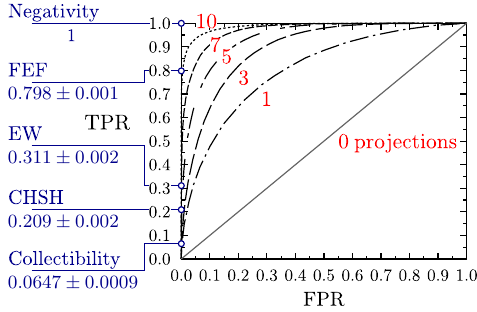}
\caption{\label{fig:two-qubits} Comparison of the ROC curves of CEW models for the two-qubit system. The vertical and horizontal axes of the graph represent the true positive rate (sensitivity) and false positive rate (fall-out), respectively. The number close to each curve indicates the length $B$ of the feature vectors used during the training process. Note that for $B = 0$ the line is diagonal due to prevalence of $0.5$. Additionally, we highlighted sensitivity values corresponding to prominent analytical entanglement measure (Negativity), Bell's inequality (CHSH) and witnesses (FEF, EW, Collectibility). It is worth noting that for the analytical entanglement measure, inequality and witnesses, the false positive rate is equal to 0.}
\end{figure}

First, we discuss the results for the two-qubit system. To evaluate the performance of our CEW models, we have evaluated a series of scenarios on a two-qubit system, as illustrated in Fig.~\ref{fig:two-qubits}. Each scenario is characterized by the number of measurement configurations $B$, which is also the length of the feature vector. The graph displays multiple ROC curves, which represent the relationship between the sensitivity (true positive rate, or TPR) and fall-out (false positive rate, or FPR) of the CEW models. Note that the $\mathrm{selectivity} = 1 - \mathrm{FPR}$.
The performance of our CEW models is quantified using the area under the curve (AUC) metric. An ideal model would achieve an AUC of 1. We have trained several CEW models using varying numbers of collective measurement configurations, as indicated by the numbers 1-10 near the ROC curves in the graph. Our results show that with a slight reduction of selectivity (up to 10\%) the sensitivity of our CEW models rises rapidly which indicates that the models are able to detect broader range of entangled systems. As expected, the highest level of performance is achieved when the ANN is provided with all available information about the investigated system, which occurs when the length of the feature vector, denoted by $B$, is equal to 10 (see Appendix for details). As the number of measurement configurations increases, resulting in AUC values are approaching 1. This demonstrates that our CEW models are effective at detecting entanglement in these systems with high accuracy. In the graph in Fig.~\ref{fig:two-qubits}, the diagonal line represents a baseline scenario in which the model has no information about the system ($B = 0$) and randomly decides, with equal probability, whether the system is entangled or separable.
\begin{figure}[t!]
\includegraphics[width=8.5cm]{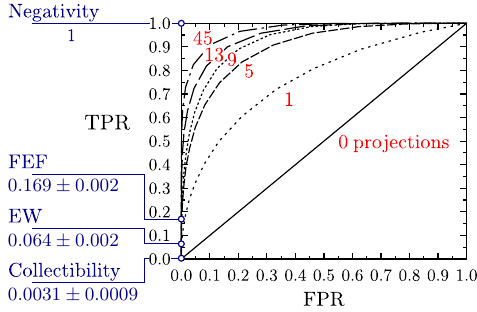}
\caption{\label{fig:qubit-qutrit} Comparison of the ROC curves of CEW models for the qubit-qutrit system. The vertical and horizontal axes of the graph represent the true positive rate (sensitivity) and false positive rate (fall-out), respectively. The number close to each curve indicates the length $B$ of the feature vectors used during the training process. Note that for $B = 0$ the line is diagonal due to prevalence of $0.5$. Additionally, we highlighted sensitivity values corresponding to prominent analytical entanglement measure (Negativity) and witnesses (FEF, EW, Collectibility). For analytical entanglement measure and witnesses, the false positive rate is equal to 0.}
\end{figure}
To provide a comparison, we also calculated the sensitivities of analytical entanglement measure (Negativity), Bell's inequality (CHSH) and witnesses (FEF, EW, Collectibility) using $1\times 10^4$ randomly sampled two-qubit systems. These sensitivities are depicted on the vertical axis of the graph.

Second, we proceed to examine the results concerning the qubit-qutrit system. Analogously to the two-qubit system, we generated multiple CEW models for this system and obtained their corresponding ROC curves, illustrated in Figure~\ref{fig:qubit-qutrit}. As expected, the performance of the CEW models improves as the number of measurement configurations increases. However, we observed diminishing returns in terms of the CEW model's performance after reaching a certain threshold, approximately at $B = 13$. Similar to our earlier findings, even slight reductions in selectivity lead to significant enhancements in the sensitivity of the CEW models. Notably, this effect becomes even more pronounced when comparing the sensitivities of our models with those of analytical entanglement measure and witnesses, where we observed a substantial reduction in sensitivities (excluding Negativity) when compared to the two-qubit system.

\paragraph*{Conclusions.}

In this study, we have successfully developed collective entanglement witnesses for two-qubit and qubit-qutrit systems using supervised learning techniques and artificial neural networks. Our aim was to analyze the CEWs sensitivity-selectivity trade-off and to improve on the range of detected entangled systems. We demonstrated the effectiveness of our CEWs by constructing the ROC curves for a given amount of input information. The ROC curves provided a comprehensive evaluation of the models' performance, showing the trade-off between sensitivity (true positive rate) and selectivity (1 - false positive rate). We observed that a slight reduction in selectivity (up to 10\%) resulted in a significant increase in sensitivity, allowing the CEW models to detect a broader range of entangled systems. Comparing the performance of our CEW models with analytical entanglement measure and witnesses, we found that our machine-designed CEWs with this slight reduction in selectivity outperformed some of the witnesses in terms of sensitivity. Note that in experimental reality, even analytical entanglement witnesses, which are designed to be perfectly selective, are prone to false positive classification due to unavoidable measurement noise. Furthermore, our approach of creating CEWs via supervised learning techniques and ANNs brings a couple of advantages. Firstly, our CEW models allow for continuous tuning of their sensitivity and selectivity, providing flexibility in entanglement detection. The trade-off between these two factors can be effectively managed by adjusting the models' decision threshold. Secondly, the concept of CEW offers high generality and requires a lower number of measurements compared to quantum state tomography. These make our method more efficient and practical for experimental implementations. Our results shed light on the fundamental relationship between the sensitivity and selectivity of entanglement detection and contribute to the development of more efficient and robust entanglement detection methods, offering new possibilities for quantum information processing tasks and quantum communication protocols. Further research can explore the extension of this approach to larger quantum systems and investigate its applicability in practical experimental scenarios.

\paragraph*{Acknowledgements.}
\begin{acknowledgments}
The authors would like to express their gratitude to Tomáš Fürst for his assistance in artificial neural networks. Authors thank Cesnet for providing data management services.
\end{acknowledgments}

\section{Appendix A: Preparation of general two-qubit and qubit-qutrit systems}
The process of generating two-qubit systems is similar to that of generating qubit-qutrit systems. First, a $4 \times 4$ ($6 \times 6$) diagonal matrix $\rho$ is randomly generated, where the diagonal elements are uniformly distributed random numbers from the range of $[0,1]$ and satisfy the condition $\mathrm{Tr}(\rho) = 1$. The next step involves creating a random unitary matrix $U$, which transforms the diagonal matrix $\rho$ into a general random density matrix $\hat{\varrho}$ using the following relation:
\begin{equation}
\hat{\varrho} = U^{\dagger}\rho U.
\end{equation}
In the last step the quantum system of two instances of $\hat{\varrho}$ is created by the relation:
\begin{equation}
\hat{\varrho}_{T} = S^{T}\hat{\varrho}S \otimes \hat{\varrho};
\end{equation}
where $S$ symbolizes a swap operator that interchanges the order of subsystems. This interchange is necessary due to the nature of collective measurements.

Note that by randomly generating $\hat{\varrho}$ the probabilities of the system being entangled or separable are not equal. For two-qubit system the probability that $\hat{\varrho}$ is separable is 63\%. In the case of qubit-qutrit system this probability is 38\%.

\section{Appendix B: Measurement configurations}

The objective is to equip the ANN with limited information regarding the analyzed system, while maintaining a high ratio between true positive rate and false positive rate. To achieve this, we utilized the concept of minimal qubit tomography for the two-qubit system. In this approach, the set of single-qubit projections consists of four projections onto states forming a tetrahedron within the Bloch sphere. One example of such a set of projections is as follows:
\begin{equation}
\begin{split}
\hat{\Pi}_1 &= \frac{1}{4}\left(\sigma_0+ \frac{1}{\sqrt{3}} \left( \sigma_1+ \sigma_2 +\sigma_3\right)\right) ,\\
\hat{\Pi}_2 &= \frac{1}{4}\left( \sigma_0+ \frac{1}{\sqrt{3}} \left( \sigma_1- \sigma_2 -\sigma_3\right)\right) ,\\
\hat{\Pi}_3 &= \frac{1}{4}\left( \sigma_0+ \frac{1}{\sqrt{3}} \left( -\sigma_1+ \sigma_2 -\sigma_3\right)\right) ,\\
\hat{\Pi}_4 &= \frac{1}{4}\left( \sigma_0+ \frac{1}{\sqrt{3}} \left( -\sigma_1- \sigma_2 +\sigma_3\right)\right) ,
\end{split}
\end{equation}
where $\sigma_{i\in\{0..3\}}$ are Pauli matrices. For two-qubit system this gives us a total of 16 combinations of local projections.

Similarly for qutrits there are single qutrit projections that form a set for minimal single qutrit tomography
\begin{equation}
\hat{\Pi}_{i} = |\psi_{i}\rangle\langle\psi_{i}|,
\end{equation} 
where $|\psi_{i\in\{1..9\}}\rangle$ are defined as
\begin{equation}
\begin{split}
|\psi_{1}\rangle &= \frac{1}{\sqrt{2}}(|0\rangle+ |2\rangle) ,\\
|\psi_{2}\rangle &= \frac{1}{\sqrt{2}}(e^{2\pi i/3}|0\rangle+e^{-2\pi i/3}|2\rangle) ,\\
|\psi_{3}\rangle &= \frac{1}{\sqrt{2}}(e^{-2\pi i/3}|0\rangle+e^{2\pi i/3}|2\rangle) ,\\
|\psi_{4}\rangle &= \frac{1}{\sqrt{2}}(|1\rangle+ |0\rangle) ,\\
|\psi_{5}\rangle &= \frac{1}{\sqrt{2}}(e^{2\pi i/3}|1\rangle+e^{-2\pi i/3}|0\rangle) ,\\
|\psi_{6}\rangle &= \frac{1}{\sqrt{2}}(e^{-2\pi i/3}|1\rangle+e^{2\pi i/3}|0\rangle) ,\\
|\psi_{7}\rangle &= \frac{1}{\sqrt{2}}(|2\rangle+ |1\rangle) ,\\
|\psi_{8}\rangle &= \frac{1}{\sqrt{2}}(e^{2\pi i/3}|2\rangle+e^{-2\pi i/3}|1\rangle) ,\\
|\psi_{9}\rangle &= \frac{1}{\sqrt{2}}(e^{-2\pi i/3}|2\rangle+e^{2\pi i/3}|1\rangle) ,\\
\end{split}
\end{equation}
with the states $|0\rangle, \hspace{1pt} |1\rangle$ and $|2\rangle$ being the basis of a single qutrit state
\begin{equation}
|0\rangle = \begin{pmatrix} 1 \\ 0 \\ 0\end{pmatrix}, |1\rangle = \begin{pmatrix} 0 \\ 1 \\ 0\end{pmatrix}, |2\rangle = \begin{pmatrix} 0 \\ 0 \\ 1\end{pmatrix}.
\end{equation}
For two qutrits, there are a total of 81 possible combinations of local projections. It is worth noting that due to the symmetry of the density matrix $\hat{\varrho}_{T}$, exchanging the local projections does not impact the result of a collective measurement. This implies that $P_{xy} = P_{yx}$. Consequently, there can be a maximum of 10 independent collective measurements for the two-qubit system, and 45 independent collective measurements for the qubit-qutrit system.

\begin{table}[h]
\caption{\label{tab:twoqubit config}List of specific measurement configurations for two-qubit system. Note that $B$ is consecutively the length of the feature vectors.}

\begin{tabular}{c}
Two-qubit system
\end{tabular}
\begin{ruledtabular} 
\begin{tabular}{ll}
$B$  & {$\text{Measurement configurations}$}                                                                                                                   \\[3pt] \hline\\[-5pt]
1 & {$\hat{\Pi}_1 \otimes \hat{\Pi}_1$}\\[3pt]

3 & {$\hat{\Pi}_1 \otimes \hat{\Pi}_1, \hat{\Pi}_2 \otimes \hat{\Pi}_2,\hat{\Pi}_3 \otimes \hat{\Pi}_3$}\\[3pt]

5  & {$\hat{\Pi}_1 \otimes \hat{\Pi}_1,\hat{\Pi}_2 \otimes \hat{\Pi}_2,\hat{\Pi}_3 \otimes \hat{\Pi}_3,\hat{\Pi}_4 \otimes \hat{\Pi}_4,\hat{\Pi}_1 \otimes \hat{\Pi}_3$}                                                           \\[3pt]

7  & {$B=5,\;\land\; \hat{\Pi}_1 \otimes \hat{\Pi}_4, \hat{\Pi}_2 \otimes \hat{\Pi}_4$} \\[3pt]

10 & {$B=7,\;\land\; \hat{\Pi}_1 \otimes \hat{\Pi}_2, \hat{\Pi}_2 \otimes \hat{\Pi}_3, \hat{\Pi}_3 \otimes \hat{\Pi}_4$}                            
\end{tabular}
\end{ruledtabular}

\begin{tabular}{c}
\\[1mm]
Qubit-qutrit system
\end{tabular}
\begin{ruledtabular} 
\begin{tabular}{ll}
$B$  & {$\text{Measurement configurations}$}                                                                                                                   \\[3pt] \hline\\[-5pt]
1 & {$\hat{\Pi}_1 \otimes \hat{\Pi}_1$}\\[3pt]

5 & {$\hat{\Pi}_1 \otimes \hat{\Pi}_1, \hat{\Pi}_3 \otimes \hat{\Pi}_3,\hat{\Pi}_5 \otimes \hat{\Pi}_5, \hat{\Pi}_8 \otimes \hat{\Pi}_8,\hat{\Pi}_9 \otimes \hat{\Pi}_9$}\\[3pt]

9  & {$B=5,\;\land\; \hat{\Pi}_2 \otimes \hat{\Pi}_2,\hat{\Pi}_4 \otimes \hat{\Pi}_4,\hat{\Pi}_6 \otimes \hat{\Pi}_6,\hat{\Pi}_7 \otimes \hat{\Pi}_7$}                                                           \\[3pt]

13  & {$B=9,\;\land\; \hat{\Pi}_1 \otimes \hat{\Pi}_2,\hat{\Pi}_3 \otimes \hat{\Pi}_4,\hat{\Pi}_4 \otimes \hat{\Pi}_5,\hat{\Pi}_8 \otimes \hat{\Pi}_9$} \\[3pt]

45 & {$\hat{\Pi}_i \otimes \hat{\Pi}_j,\;i,j\in[1,9],j\geq i$ }                            
\end{tabular}
\end{ruledtabular}
\end{table}

\end{document}